\documentclass{article}

\usepackage{PRIMEarxiv}

\usepackage[utf8]{inputenc} 
\usepackage[T1]{fontenc}    
\usepackage{hyperref}       
\usepackage{url}            
\usepackage{booktabs}       
\usepackage{amsfonts}       
\usepackage{nicefrac}       
\usepackage{microtype}      
\usepackage{lipsum}
\usepackage{fancyhdr}       
\usepackage{graphicx}       
\graphicspath{{images/}}     

\usepackage{cite}
\usepackage{amsmath,amssymb,amsfonts}
\usepackage{algorithm,algorithmic}

\usepackage{makecell}

\title{Binary Neural Network Implementation for Handwritten Digit Recognition on FPGA}

\author{
  Emir Devlet Ertörer \\
  Faculty of Engineering and Natural Sciences\\ 
  Department of Computer Engineering \\
  Yeditepe University \\
  Istanbul, Türkiye\\
  \texttt{emirdevlet.ertorer@std.yeditepe.edu.tr} \\
     \And
  Cem Ünsalan\\
 Faculty of Engineering and Natural Sciences\\ 
 Department of Computer Engineering \\
  Yeditepe University \\
  Istanbul, Türkiye\\
  \texttt{unsalan@yeditepe.edu.tr} \\
}

\begin{document}
\maketitle

\begin{abstract}
Binary neural networks provide a promising solution for low-power, high-speed inference by replacing expensive floating-point operations with bitwise logic. This makes them well-suited for deployment on resource-constrained platforms such as FPGAs. In this study, we present a fully custom BNN inference accelerator for handwritten digit recognition, implemented entirely in Verilog without the use of high-level synthesis tools. The design targets the Xilinx Artix-7 FPGA and achieves real-time classification at 80\,MHz with low power consumption and predictable timing. Simulation results demonstrate 84\% accuracy on the MNIST test set and highlight the advantages of manual HDL design for transparent, efficient, and flexible BNN deployment in embedded systems. The complete project including training scripts and Verilog source code are available at \cite{bnn_repo} for reproducibility and future development.
\end{abstract}

    
\section{Introduction}

The progress in the field of deep learning has led to major advancements in speech recognition, generative AI, large language models, and computer vision. The widely used neural network models used in these applications have achieved state of the art performance across a wide range of tasks. Despite their high accuracy, the models rely on floating-point computations that typically use 32-bit representations. Moreover, they have large number of parameters to be set ranging from millions to billions. These lead to significant computational cost and memory usage. Therefore, several model compression techniques have been proposed for this purpose. They try to reduce the memory and computational resource usage without severely impacting performance and accuracy. 

One of the most effective compression techniques used for this purpose is quantization. Here, weight and activation values in a neuron are represented using low-precision numbers. Binary quantization is the most extreme form of quantization in this category. It restricts weights and activations to two possible values as -1 and +1 or 0 and 1. This form of quantization creates special class of models known as binary neural networks (BNNs). BNNs significantly reduce memory usage by representing both weights and activations using only 1 bit instead of 32 bits for floating-point values. Moreover, expensive floating-point matrix multiplications in neural network computations can be replaced with lightweight bitwise operations such as XNOR and popcount, which offer significant speedup. For instance, BNNs can achieve up to 32 times memory savings and up to 58 times computational speedup on CPUs \cite{qin2020}. 

 We can summarize work in literature on BNNs as follows. Courbariaux~\emph{et al.}~\cite{courbariaux2016} introduced the BinaryNet, which used the sign function for binarization along with a straight-through estimator (STE) to enable gradient propagation during training. Rastegari~\emph{et al.}~\cite{rastegari2016} extended this approach with XNOR-Net, introducing scaling factors to improve accuracy in convolutional architectures for large-scale tasks such as ImageNet classification. FPGAs are a natural fit for BNN acceleration due to their reconfigurable fabric, inherent parallelism, and potential for custom hardware optimization. Their configurable logic and on-chip memory make them efficient platforms for implementing bitwise operations. Therefore, several frameworks targeted BNN deployment on FPGAs using high-level synthesis (HLS) tools. Umuroğlu~\emph{et al.}~\cite{umuroglu2017} proposed FINN, a dataflow-style framework composed of matrix-vector units for scalable BNN inference on reconfigurable hardware. Zhao~\emph{et al.}~\cite{zhao2017} developed a system for accelerating binary convolutional networks on programmable FPGAs using software-defined techniques. Prior studies have shown that BNNs can achieve competitive accuracy on benchmarks like MNIST and CIFAR-10 while offering substantial gains in energy efficiency and runtime performance \cite{hubara2018}. The reader can consult review papers on BNNs for a comprehensive overview of their current status and developments \cite{zhao2020, qin2020, sayed2023}.

While the previous methods have significantly advanced BNN deployment on hardware, they largely focus on convolutional architectures and relied on HLS tools for hardware generation or frameworks that abstract the underlying logic. While these tools accelerate development, they limit visibility into how the network operates at a low level, especially in terms of data flow, timing and logic decisions. This lack of transparency makes it harder to optimize, debug or fully understand the hardware behavior of the systems.

We propose a fully connected BNN inference system from scratch in Verilog without any use of HLS tools to overcome the shortcomings in previous studies. The target is handwritten digit classification using the MNIST dataset. All inference steps like binarized weight loading, XNOR-popcount operations, threshold comparison and argmax output are implemented using cycle-accurate FSMs on a Nexys A7-100T FPGA. This allows full control over memory layout, logic structure and timing which provides a transparent framework for understanding binary inference at hardware level. Therefore, this implementation provides direct insight into how each bit is processed, how intermediate values are handled and how control flows between layers. No other BNN implementation has been found that provides this level of manual control over both the model architecture and its bitwise execution on FPGA. The study focuses on the FPGA based inference stage of a trained BNN. The model is trained offline using Python and TensorFlow and then exported in binary format for hardware deployment. The design has a few limitations. First, it only supports fully connected architectures and convolutional layers are not included. Second, test images, weights and thresholds are stored in static ROM, which means changing the input set or the network architecture requires re-synthesis. Third, the Verilog code is specific to the trained model architecture because the FSM and memory accesses are hardcoded for a particular number of layers and neurons. Supporting a different model would require significant code changes as each layer is implemented as a separate FSM state with fixed dimensions.

\section{Fundamental Operations in BNNs}

We introduce fundamental BNN operations in this section before explaining the proposed method. Our aim is twofold. First, we would like to get familiar with BNN operations. Second, we would like to show possible extension of our approach to other BNN calculations. Therefore, we will cover forward propagation, backpropagation, batch normalization, activation functions, and inference. To emphasize again, our main focus will be inference in this study.

\subsection{Forward Propagation}

To understand how forward propagation works in BNNs, it is important to first understand how standard neural networks work. In any neural network, each layer contains a set of weights (learnable parameters) that determine how much influence each input has on the output of a neuron. These weights are initially assigned random values and are updated through training so that the network can learn patterns from the provided data.

During forward propagation, each neuron receives a set of input values, called activations, and computes a weighted sum of these inputs, as $z = \sum_{i=1}^{n} w_i \cdot x_i$. The value \(z\) is then passed through an activation function (such as the ReLU or sigmoid function) to produce the neuron output. This step helps the network to respond the input in a meaningful way. The weighted sum shows how much the input matches what the neuron has learned to look for. The activation function then decides how the neuron should react to this input. By doing this selectively, the network can learn patterns and classify different outputs or classes.

BNNs simplify this process by constraining both weights \(w_i\) and activations \(x_i\) to binary values. This allows floating-point multiplications to be replaced by bitwise XNOR operation. In binary logic, the XNOR gate outputs 1 when both inputs are the same and 0 otherwise. Applying the XNOR operation to each element in two binary vectors \(x\) and \(w\), we obtain a new binary vector that shows which elements match. The number of ones in the result, called the popcount, tells us how many inputs matched their corresponding weights, as $\text{popcount}(\text{XNOR}(x, w)) = \sum_{i=1}^n [x_i = w_i]$. This count by itself does not represent the full dot product. In BNNs, the idea is to simulate the dot product using only bitwise logic. When \(x_i\) and \(w_i\) are the same (both +1 or both -1), their product is +1. If they differ, the product is -1. Hence, each match adds 1 and each mismatch subtracts 1 from the sum.

Let \(m\) be the number of matching elements and \(n\) be the total number of elements in the vectors. Then the number of mismatches is \(n - m\) and the signed dot product becomes $z = (m \cdot (+1)) + ((n - m) \cdot (-1))$. This can be simplified as $z = 2m - n$. Since \(m = \text{popcount}(\text{XNOR}(x, w))\), we can express the full dot product approximation as $z \approx 2 \cdot \text{popcount}(\text{XNOR}(x, w)) - n$. This transformation allows us to simulate the original full dot product between binary vectors using only XNOR and popcount operations, which are more efficient in hardware than multiplications and additions.

Both the input activations and weights should first be binarized to make the XNOR based computation possible. This is done using the sign function which converts real valued inputs into binary form, as

\begin{equation}
\text{sign}(z) = 
    \begin{cases}
        +1, & z \geq 0 \\
        -1, & z < 0
    \end{cases}
\label{eq:sign_func}
\end{equation}

\noindent The network converts values into binary form by applying the sign function in the forward pass. This binarization allows replacing standard multiplications with faster XNOR and popcount operations. Once binary activations are computed, they are compared to true labels to calculate the loss. This loss is then used to adjust weights during backpropagation.

\subsection{Backpropagation}

To understand how backpropagation works in BNNs, it is important to first understand the standard operation. In a typical neural network, once the forward pass produces predictions, the network calculates a loss to measure how different predictions are from actual labels. This loss is then sent backward through the network using the chain rule so that the weights can be updated to reduce future errors. This process works well for differentiable activation functions like ReLU or sigmoid.

Backpropagation in BNNs faces a major problem due to the use of the sign function during binarization. The sign function is non-differentiable at zero and its derivative is zero almost everywhere else. As a result, the gradient is either undefined or it vanishes, which makes it impossible to propagate error signals backward. This causes what is known as the vanishing gradient problem where gradients become too small for the network to keep learning. In BNNs, this issue is even more serious since the sign function does not give any meaningful gradient during backpropagation. Without a solution, the model cannot update its weights properly, which stops learning altogether \cite{bengio2013}. To solve this issue, BNNs use a method called the straight-through estimator (STE). Instead of relying on the actual derivative of the sign function that does not exist, STE treats the sign function as if it is the identity function during backpropagation. This way, it skips over the non-differentiable binarization step and assigns an approximate gradient to keep the learning going.

The STE approximation is typically written as

\begin{equation}
\frac{d}{dx} \text{sign}(x) \approx 
\begin{cases}
    +1, & |x| \leq 1 \\
    -1, & \text{otherwise}
\end{cases}
\label{eq:ste_derivative}
\end{equation}

\noindent This means that if the input is within a certain range, the gradient is approximated as one. Outside that range, it is taken as zero. This keeps gradients flowing during training and allows standard optimizers like Adam or AdaMax to work effectively. This methods allows the network keep updating its weights even though it does not use the real gradient from the binarization step. While the exact size of the gradient might not be accurate, STE keeps the direction of the update which is often good enough for the model to learn. Thanks to this property, BNNs can still be trained using standard optimization methods, even though they use only binary weights and activations during the forward pass.

\subsection{Batch Normalization and Activation Functions}

The sign function is used to binarize both weights and activations in BNNs. Since this function is very sensitive to inputs near zero, even a small change in the input can flip the sign and change the binary result. This makes it very important to carefully control the distribution of inputs before binarization to ensure stable and meaningful outputs across layers. To address this, batch normalization is proposed by Ioffe and Szegedy~\cite{ioffe2015}. This operation is used right before each binarization step. Batch normalization adjusts inputs such that they have mean zero and variance one across the mini-batch. This keeps the values in a predictable range and makes the binarization step more stable and effective during training.

The batch normalization operation is defined as

\begin{equation}
    \text{BN}(x) = \gamma \cdot \frac{x - \mu_B}{\sqrt{\sigma_B^2 + \epsilon}} + \beta
    \label{eq:batchnorm_def}
\end{equation}

\noindent where \( \mu_B \) and \( \sigma_B^2 \) represent the mean and variance of the input \(x\) over the batch, respectively. The terms \( \gamma \) and \( \beta \) are learnable parameters that allow the model to rescale and shift the normalized output. The small constant \( \epsilon \) is added to prevent division by zero. The normalization is typically simplified by fixing \( \gamma = 1 \), \( \beta = 0 \) and disabling scaling during inference in BNN implementations. Once the input is normalized, it is binarized using the sign function applied to the batch-normalized output.

In BNNs, batch normalization and binarization replaces traditional activation functions like ReLU. Instead of applying a separate non-linear function, the network takes the sign of the normalized activation to produce a binary output for the next layer. It helps keeping the activation values centered around zero and makes binary outputs more balanced. This prevents neurons from always producing the same value and improves gradient flow during training which allows the model to learn more efficiently.

\subsection{Inference}

The BNN does not require gradient computations or weight updates during inference. All hidden layers operate entirely with binarized weights and activations, while the output layer retains full-precision activations when probabilistic outputs or external interface compatibility is needed. This binarized structure enables the entire computation to be reduced to efficient bitwise operations, making BNNs highly suitable for resource-constrained hardware implementations. The following high-level pseudocode outlines the logic of the binary forward inference procedure used in this study.

\begin{algorithm}[htbp]
\caption{BNN inference algorithm}
\begin{algorithmic}[1]
\renewcommand{\algorithmicrequire}{\textbf{Input:}}
\renewcommand{\algorithmicensure}{\textbf{Output:}}
\REQUIRE Binary input vector, weight and threshold matrices
\ENSURE Final binary output activations
\STATE Initialize \texttt{input\_vector} as binary input
\FOR{each layer $l = 1$ to $L$}
    \STATE Load binary weights $W_l$ and thresholds $T_l$
    \FOR{each output neuron $j$ in layer $l$}
        \STATE Set \texttt{popcount} $\leftarrow 0$
        \FOR{each input bit $i$}
            \STATE $x_i \leftarrow \texttt{input\_vector}[i]$
            \STATE $w_{ij} \leftarrow W_l[i][j]$
            \IF{$x_i$ XNOR $w_{ij} = 1$}
                \STATE Increment \texttt{popcount}
            \ENDIF
        \ENDFOR
        \STATE $z \leftarrow 2 \times \texttt{popcount} - N$
        \IF{$z \geq T_l[j]$}
            \STATE Append 1 to \texttt{next\_activations}
        \ELSE
            \STATE Append 0 to \texttt{next\_activations}
        \ENDIF
    \ENDFOR
    \STATE \texttt{input\_vector} $\leftarrow$ \texttt{next\_activations}
\ENDFOR
\RETURN \texttt{input\_vector}
\end{algorithmic}
\end{algorithm}

\section{The Proposed Method}

We summarize the main steps of the proposed method in this section. Our design integrates model architecture, training, and hardware implementation through bitstream generation, enabling low-power, low-latency digit recognition on the MNIST dataset. All logic was implemented in Verilog without HLS tools to enable cycle-accurate control of computation and dataflow.

\subsection{Model Architecture and Training}

We train the BNN model using the TensorFlow framework with Keras as its high-level API. While doing so, we use the quantization-aware training available in TensorFlow. We use the Larq library which extends Keras by providing binary compatible layers to support binarized operations. We trained the BNN using the MNIST dataset, which comprises 60000 training and 10000 test images of size 28$\times$28 pixels. Each image is flattened to a 784-dimensional input vector and normalized to [-1,1]. Therefore, the network architecture includes an input layer of 784 features. The first hidden layer has 128 neurons with binarized weights and activations, using sign activation and batch normalization. The second hidden layer has 64 neurons with binarized weights and activations, also using sign activation and batch normalization. The output layer has 10 neurons with binarized weights and real-valued activations to enable argmax classification. Batch normalization layers are placed after each hidden and output layer to improve convergence and stability. This ensures consistent feature scaling and supports threshold folding for FPGA inference.

We trained the model using the Adam optimizer with sparse categorical cross-entropy loss, batch size of 64, and quantization-aware training for 15 epochs. The exponential learning rate decay schedule is applied to improve training stability and convergence. The learning rate starts at 0.001 and decays by a factor of 0.96 every 1000 steps using a staircase strategy, which means that the rate remains constant within each interval before dropping at fixed steps. These hyperparameters were set experimentally. Higher starting rates like 0.01 led to unstable training and sharp fluctuations in accuracy while lower values like 0.0005 resulted in slow convergence. The selected configuration provided the best balance in achieving smooth and fast convergence without overfitting or divergence.

To minimize resource usage on the FPGA, batch normalization layers were folded into neuron-specific threshold values, eliminating the need for explicit normalization logic during inference \cite{ioffe2015}. The folded threshold $\theta$ for each neuron is computed as

\begin{equation}
\theta = \left\lfloor \beta - \frac{\mu}{\sqrt{\sigma^2 + \epsilon}} \right\rceil
\end{equation}

\noindent where $\beta$ is the batch normalization bias, $\mu$ is the moving mean, $\sigma^2$ is the moving variance, and $\epsilon$ is a small constant for numerical stability. These thresholds were quantized as 11-bit signed integers and saved in binary \texttt{.mem} files, alongside the binarized weight matrices. To match the FPGA's access patterns, exported weights are transposed such that each ROM row corresponds to a full set of input weights for a single neuron. This enables efficient parallel access during inference. The model is then saved in the \texttt{.h5} format containing both the network weights and batch normalization statistics. This format made it easy to extract and convert data for FPGA implementation.


\subsection{Model Export and Hardware Formatting}

After training, weights and batch normalization parameters are extracted using paths corresponding to each \texttt{QuantDense} and \texttt{BatchNormalization} layer from the \texttt{.h5} file that stored the trained model. Since the model uses no bias terms, only kernel weights and normalization statistics (moving mean, variance and beta offset) are extracted. These values are later converted to binary form or threshold representations and exported as \texttt{.mem} files. To match the memory access pattern of the FPGA, weight matrices are transposed after exporting them. This ensures that each row in the memory corresponds to a single neuron’s full set of weights to enable efficient ROM access during inference.

Thresholds were extracted and folded for the first two hidden layers to allow the XNOR–popcount outputs to be compared directly against them using simple comparators. However for the output layer, raw floating-point logits were kept instead of binarizing the outputs. This was done to preserve classification accuracy, as the final activation outputs are more sensitive to quantization and directly influence the predicted class.

\subsection{Hardware Architecture}

Our design follows standard digital system design principles to ensure predictable behavior, efficient timing, and modular structure across all components \cite{unsalan2017}. It consists of dual-port BRAMs, which are chosen for weight storage due to their high density and dual-port capability, enabling parallel weight fetches for XNOR operations. LUT-based ROMs are used for thresholds to minimize BRAM usage while providing fast access to folded batch normalization parameters. A centralized finite state machine (FSM) controls sequential layer computation, popcount accumulation, threshold comparison, and argmax classification. Finally, a seven-segment display decoder converts the predicted digit into display signals.

\subsection{Finite State Machine Control}

The inference process is coordinated by the FSM that operates through five sequential stages: first hidden layer computation, second hidden layer computation, output layer computation, output classification, and completion. The FSM begins by processing the input layer, where it performs XNOR operations between input bits and corresponding weights, accumulates match counts, computes dot product approximations, and applies threshold comparisons to generate binary activations for the first hidden layer. It then advances to the second hidden layer, following a similar sequence of operations with adjusted accumulation and thresholding parameters appropriate for that layer. In the output layer stage, XNOR operations and accumulation are performed, but no thresholding is applied. Hence, raw sums are retained for final classification. The classification stage identifies the class index with the highest output score through iterative comparison. The final stage signals the completion of the inference process and holds the result steady until the system is reset. Internal counters, control flags, and synchronization signals are used throughout the operation to ensure correct data flow, precise state transitions, and consistent layer-by-layer execution.

\subsection{Implementation Details}

Scalability in the design is achieved through a configurable parameter that defines the number of neurons processed in parallel during inference. This parameter allows the architecture to balance inference speed against hardware resource utilization. The weight and threshold memories are instantiated using parameterized generate loops, ensuring that both the memory structure and indexing logic dynamically adapt to the chosen parallelism level. The design consumes more BRAM as parallelism increases. The synthesis tool automatically maps remaining ROM structures onto LUTs to meet memory requirements when BRAM resources are exhausted. The system operates at a clock frequency of 80\,MHz, which is selected to provide a practical balance between reliable timing closure and sufficient throughput for real-time embedded applications. This frequency is verified during synthesis and implementation to achieve timing closure without setup or hold violations while maintaining thermal and power margins suitable for low-power operation.

\subsection{Resource Utilization and Power}

Post-implementation reports indicate that the design utilized 26.23\% of the LUTs and 10.42\% of the flip-flops available on the Artix-7 XC7A100T FPGA at a parallelism level of 64 neurons. This reflects the logic required for XNOR-popcount operations, FSM control, and address generation. BRAM usage reached 97.78\%, with 132 out of 135 BRAM blocks consumed. This high usage was expected due to the instantiation of large dual-port ROMs needed for weight storage across layers and the additional memory required to support wide parallel neuron access. I/O utilization was modest at 6.67\%, allocated for clocking, reset, seven-segment display interfacing, and debugging signals. Power analysis estimated total on-chip power at approximately 0.617\,W, with dynamic power contributing 83\% of this value. The dominant source of dynamic consumption was BRAM activity, which accounted for 74\% of the dynamic power. The remaining contributions came from clocks (11\%), signal nets (8\%), logic elements (8\%), and negligible I/O switching. The junction temperature was reported as 27.8$^\circ$C, well within safe thermal margins for reliable operation. While power consumption was relatively high for this class of FPGA, it reflected a realistic memory-intensive architecture designed to balance throughput and resource feasibility.

\section{Experimental Results}

We evaluated the proposed design through functional simulation and post-implementation analysis. The evaluation covered correctness verification, speed versus resource trade-off analysis, BNN versus CNN performance comparison, platform comparison results, timing closure checks, and power estimation. These tests were conducted to validate functional accuracy, measure performance, and assess scalability and efficiency.

\subsection{Correctness Verification}

We verified the correctness using 100 binarized MNIST test images, with 10 representative samples for each digit from 0 to 9. Each image is normalized, binarized, converted into a \texttt{.mem} file, and loaded into the FPGA design during simulation. The predicted digit was recorded and compared to true labels for each case. Our hardware implementation achieved 84\% accuracy, correctly classifying 84 out of 100 test cases. This outcome is consistent with the 87.97\% accuracy of the trained model on the full MNIST test set, confirming that the FPGA design replicates the intended inference behavior. 

\subsection{Parallelism vs Speed and Resource Usage}

We synthesized and implemented the proposed design across a range of parallelization levels to evaluate the trade-off between inference speed and resource usage. This comparison assessed how memory style impacts latency, logic utilization, power consumption, and scalability as parallelism increases. We changed the parallelism level from 1 to 128 neurons processed per cycle. We tabulate the results in Table~\ref{tab:speed_resource}.

\begin{table}[htbp]
\caption{Latency, speedup, and resource usage for various parallelism and memory styles.}\label{tab:speed_resource}
\begin{center}
\begin{tabular}{rrrrrrrcc}
\hline
\textbf{Parallelism} & \textbf{Latency} & \textbf{Speedup} & \textbf{LUTs} & \textbf{Flip-Flops} & \textbf{BRAMs} & \textbf{Power} & \textbf{Dynamic/Static} & \textbf{Memory} \\
                                     & \textbf{(ns)}            &                                  & \textbf{(\%)} & \textbf{(\%)} & \textbf{(\%)} & \textbf{(W)} & \textbf{(\%)} & \textbf{Style} \\
\hline
1   & 1{,}096{,}045 & 1.00 & 1.24 & 0.36 & 9.63 & 0.103 & 5/95 & BRAM \\
1   & 1{,}096{,}035 & 1.00 & 3.92 & 0.38 & 0 & 0.106 & 9/91 & LUT \\
4   & 274{,}465 & 4.00 & 2.62 & 0.39 & 38.52 & 0.111 & 10/90 & BRAM \\
4   & 274{,}455 & 4.00 & 10.49 & 0.53 & 0 & 0.119 & 19/81 & LUT \\
8   & 137{,}645 & 7.96 & 4.88 & 0.48 & 77.04 & 0.127 & 20/80 & BRAM \\
8   & 137{,}635 & 7.96 & 20.43 & 0.61 & 0 & 0.115 & 16/84 & LUT \\
16  & 68{,}905 & 15.90 & 16.35 & 4.51 & 97.78 & 0.183 & 43/57 & BRAM \\
16  & 68{,}895 & 15.90 & 21.74 & 0.78 & 0 & 0.142 & 32/68 & LUT \\
32  & 34{,}865 & 31.43 & 22.71 & 12.53 & 97.78 & 0.633 & 83/17 & BRAM \\
32  & 34{,}855 & 31.45 & 18.20 & 0.96 & 0 & 0.147 & 34/66 & LUT \\
64  & 17{,}845 & 61.42 & 26.02 & 8.41 & 97.78 & 0.617 & 83/17 & BRAM \\
64  & 17{,}835 & 61.45 & 24.09 & 1.46 & 0 & 0.156 & 37/63 & LUT \\
128 & 9{,}865 & 111.10 & 29.38 & 2.48 & 0 & 0.179 & 46/54 & LUT \\
\hline
\end{tabular}
\end{center}
\end{table}

\subsubsection{Latency Comparison}

As can be seen in Table~\ref{tab:speed_resource}, increasing parallelism substantially reduced inference latency. This improvement came at the cost of high BRAM usage, reaching 97.78\% at 64 neurons per cycle with BRAM weight storage. The LUT-based configuration avoided BRAM use but required more logic resources in lower parallelization levels. Dynamic power consumption increased with parallelism due to higher switching activity. Junction temperature remained within safe operating margins across all configurations. BRAM-based design failed to synthesize due to resource limits and lack of LUT fallback beyond the 64 level parallelism. The 128-level configuration used fully LUT-based memory, with higher levels proving unsynthesizable, marking 128 as the practical upper limit for our design.

The BRAM and LUT based versions show nearly identical latency values across all parallelization levels up to 64. The observed difference between the two memory styles is only 10\,ns, indicating that from a pure timing perspective both styles perform similarly. This reflects the fact that the control logic was designed to process neurons one at a time although dual-port BRAM provides two weight rows per cycle and could support processing two neurons in the current layer at once. This design choice kept the logic simple and easier to verify but it meant that the available memory bandwidth was not fully utilized which limits the latency improvement that dual-port BRAM could offer.

\subsubsection{Speedup Comparison}

As can be seen in Table~\ref{tab:speed_resource}, the speedup is not perfectly linear while the inference latency decreases with each doubling of parallelism. For instance, the speedup is 15.9x instead of the ideal 16x at 16x parallelism. At 64x parallelism it drops to 61.4x instead of the ideal 64. This difference becomes more noticeable at 128x parallelism where the speedup reaches only 111.1x compared to the expected 128x. The nonlinearity in speedup results from several hardware factors. As the parallelism increases, more computational units are instantiated which puts additional stress on routing resources. This leads to longer interconnect delays and less optimal placement, reducing timing efficiency. Moreover, memory contention increases because more parallel units try to read from memory at the same time. This creates access bottlenecks even with dual-port BRAM or LUT based ROMs. Control overhead also increases with more FSM logic and wider popcount logic, since more bits need to be counted simultaneously in each clock cycle. This adds delay and limits speedup. These effects cause diminishing returns as parallelism reaches higher levels.

\subsubsection{BRAM and LUT Usage}

Table~\ref{tab:speed_resource} summarizes post-implementation resource usage across increasing levels of parallelism. Each value is accompanied by its percentage relative to the total available resources on the Xilinx Artix XC7A100T FPGA. BRAM utilization increases rapidly up to 16x parallelism and reaches 132 blocks, which is 97.78\% of the Xilinx Artix XC7A100T’s BRAM capacity. Beyond 64x parallelism, the BRAM based design cannot synthesize at higher parallelism levels as it no longer supports partial fallback to LUTs. Hence, the 128x configuration was tested using a fully LUT based implementation. Attempts to synthesize even higher parallel configurations with LUT only memory also failed, marking 128x as the practical upper limit for parallelization in this design.

LUT usage in BRAM based configurations grows steadily with parallelism, starting from only 1.24\% at 1x and increasing up to 26.02\% at 64x. LUT based configurations begin with significantly higher LUT usage, 3.92\% even at 1x due to logic overhead from synthesizing ROMs out of LUTs. While the LUT usage continues to grow in BRAM based designs with each level of parallelism, the LUT only designs show a different behavior where LUT usage rises until 16x parallelism (21.74\%) but then drops at 32x to 18.20\% before increasing again. This drop probably reflects optimization effects in synthesis such as logic folding or reduced duplication in control logic when large ROMs dominate the design.

At 32x and 64x levels, the BRAM based designs actually use more LUTs than their LUT only equivalents. This is likely due to the added complexity of managing many parallel dual-port BRAM instances. Each BRAM block requires dedicated address generation logic, read-enable control and synchronization signals, all of which are implemented using LUTs. As the number of parallel accesses increases, this overhead also increases, leading to a sharp increase in control and routing logic. In comparison, the LUT only versions benefit from a more compact memory organization. Since the weights are synthesized directly into the logic fabric as distributed ROMs, address decoding and access logic can be flattened and optimized during synthesis. This eliminates the need for deep FSM coordination and shared BRAM access control, allowing the tools to simplify the datapath. 

\subsubsection{Flip-Flop Usage}

As can be seen in Table~\ref{tab:speed_resource} flip-flop (FF) usage follows a similarly distinct pattern. In BRAM based configurations, FF usage increases significantly at higher parallelism levels, peaking at 12.53\% for 32x, due to deeper sequential control structures and wider data movement. FF usage drops to 8.41\% at 64x despite higher parallelism. This reduction might be due to architectural optimizations in the control FSMs or shared datapaths that reduce duplication or potential logic folding applied by the synthesis tool to meet timing constraints under increased routing pressure. However, FF utilization remains consistently low, reaching only 1.46\% at 64x and 2.48\% at 128x in LUT based designs. This suggests that the logic structure of LUT based memory may require fewer sequential elements for control and buffering or certain datapath and control logic is implemented as combinational rather than registered logic due to synthesis optimizations. Overall, the table demonstrates that while BRAM becomes a limiting factor early in the scaling process, logic resources remain comfortably within the device's limit. 

\subsubsection{The Dynamic vs Static Power Ratio}

The dynamic vs static power ratio reveals how activity changes with parallelism. As can be seen in Table~\ref{tab:speed_resource}, most power is static with dynamic power around 5–20\% at low levels. As more units become active, dynamic power rises due to increased switching especially in BRAM designs where it exceeds 80\% at 16–64x. This suggests that avoiding BRAM reduces high activity memory access and keeps dynamic power lower.

Overall, BRAM based designs are more power hungry at higher configurations due to intensive memory access and wide datapath control while LUT only designs trade off slightly higher LUT usage for better energy efficiency. This also reflects in thermal behavior where the LUT only designs maintains lower and more stable junction temperatures. Thus, while BRAM configurations deliver higher throughput, they come with higher power and thermal costs.

\subsection{Timing Slack}

Table~\ref{table:timing_slack_comparison} shows the worst negative slack (WNS) and worst hold slack (WHS) reported by Vivado after place-and-route for each configuration. WNS reflects how close the design is to violating setup timing constraints. Lower values mean less time left before an error occurs. WHS shows if signals are arriving too early which can cause data to change when it should not during a clock cycle.

\begin{table}[htbp]
\caption{Post Implementation timing slack values across configuration at 80\,MHz.}\label{table:timing_slack_comparison}
\centering
\begin{tabular}{cccc}
    \hline
    \textbf{Parallelization} & \makecell{\textbf{WNS} \\ \textbf{(ns)}} & \makecell{\textbf{WHS} \\ \textbf{(ns)}} & \makecell{\textbf{Memory} \\ \textbf{Style}} \\
    \hline
    1   & 1.144 & 0.169 & BRAM \\
    1   & 3.564 & 0.115 & LUT \\
    4   & 1.525 & 0.132 & BRAM \\
    4   & 1.975 & 0.039 & LUT \\
    8   & 1.043 & 0.062 & BRAM \\
    8   & 1.708 & 0.187 & LUT \\
    16  & 0.370 & 0.033 & BRAM \\
    16  & 1.109 & 0.050 & LUT \\
    32  & 0.680 & 0.075 & BRAM \\
    32  & 1.950 & 0.129 & LUT \\
    64  & 0.939 & 0.081 & BRAM \\
    64  & 0.519 & 0.040 & LUT \\
    128 & 1.163 & 0.025 & LUT \\
    \hline
\end{tabular}
\end{table}

As can be seen in Table~\ref{table:timing_slack_comparison}, WNS generally drops with higher parallelism due to added logic and routing delays across both memory styles. In BRAM designs, WNS drops to 0.37\,ns at 16x but slightly improves at 32x and 64x probably from better placement or reduced congestion in those builds. For LUT based designs, WNS starts high at 1x (3.564\,ns), stays above 1\,ns up to 32x, goes below 1 at 64x due to increased complexity, then improves at 128x which may be because the fully flattened ROM layout reduces long timing paths. WHS values remain small across all configurations, ranging from 0.025\,ns to 0.187\,ns. However, they always stay positive. Overall all configurations meet the 80\,MHz timing target. LUT-only designs show better WNS in mid-range configurations like 16x and 32x while BRAM based designs generally have more consistent hold slack performance.

\subsection{Power and Thermal Estimates}

Table~\ref{table:thermal_comparison} indicates that power and temperature values increase with parallelization. However, the rate of increase is different between memory types. BRAM designs have steeper power rise while LUT versions grow more gradually and stay efficient.

\begin{table}[htbp]
\caption{Post Implementation power and temperature estimates across configurations.}\label{table:thermal_comparison}
\centering
\begin{tabular}{ccccc}
\hline
\textbf{Parallelization} & \makecell{\textbf{Total} \\ \textbf{Power (W)}} & \makecell{\textbf{Junction} \\ \textbf{Temperature (°C)}} & \makecell{\textbf{Dynamic/Static} \\ \textbf{Power (\%)}} & \textbf{Memory Style} \\
    \hline
    1   & 0.103 & 25.5 & 5/95 & BRAM \\
    1   & 0.106 & 25.5 & 9/91 & LUT \\
    4   & 0.111 & 25.5 & 10/90 & BRAM \\
    4   & 0.119 & 25.5 & 19/81 & LUT \\
    8   & 0.127 & 25.6 & 20/80 & BRAM \\
    8   & 0.115 & 25.5 & 16/84 & LUT \\
    16  & 0.183 & 25.8 & 43/57 & BRAM \\
    16  & 0.142 & 25.6 & 32/68 & LUT \\
    32  & 0.633 & 27.9 & 83/17 & BRAM \\
    32  & 0.147 & 25.7 & 34/66 & LUT \\
    64  & 0.617 & 27.8 & 83/17 & BRAM \\
    64  & 0.156 & 25.7 & 37/63 & LUT \\
    128 & 0.179 & 25.8 & 46/54 & LUT \\
    \hline
\end{tabular}
\end{table}

Junction temperature represents the internal temperature of the internal chip, which is influenced by dynamic power consumption and the efficiency of thermal dissipation. As can be seen in Table~\ref{table:thermal_comparison}, it increases as the design consumes more power and utilizes more logic resources. For BRAM designs, the temperature rises from 25.5\,°C at 1x to a peak of 27.9\,°C at 32x thats reflected by the jump in power from 0.103\,W to 0.633\,W. LUT based designs maintain nearly constant temperatures around 25.5–25.8\,°C across all levels that is reflected by their lower total power consumption.

\subsection{Summary and Trade-Off Justification}

Summaries and detailed analysis in previous subsections show the key trade-offs between performance, resource usage and efficiency. Both BRAM and LUT based memory styles differ significantly at higher configurations. The 64x BRAM based implementation is selected as the final configuration since it provides the best balance between performance, hardware feasibility and realistic deployment conditions. It delivers an inference latency of 17,845 ns, offering a 61.4x speedup over the baseline while fully utilizing the available 132 BRAM blocks (97.78\%). Logic utilization remains  moderate (26.02\% LUTs, 8.41\% FFs) and the design meets timing at 80\,MHz with 0.939\,ns of WNS. Although its power consumption is relatively high at 0.617\,W, this is compensated by its resource realism.

While the 128x LUT only design performs correctly in simulation and consumes significantly less power (0.179\,W), its practicality is limited. It achieves only a 1.8x speedup over the 64x design while requiring the highest LUT usage (29.38\%) and offers no further scalability as synthesis fails beyond this point. Relying only on LUT based ROMs to store model weights is not representative of real FPGA deployment scenarios. In practice, model weights are stored in larger, external or on-chip SRAM blocks or mapped to dedicated BRAM when available. Using LUTs to synthesize memory for large models is inefficient and not scalable. Moreover, BRAM based memory access more accurately reflects hardware implementations where high throughput and low latency access to structured memory is very important. The LUT only implementation would become impractical for larger networks where weights can span millions of parameters even though it is functional in this controlled design. Therefore, the BRAM configuration is a better approximation for a realistic memory hierarchy and it is a more meaningful platform for analyzing power, timing and utilization behaviors under real-world conditions. In summary, the 64x BRAM based configuration achieves the optimal trade-off: it maximizes parallelism without exceeding physical limits, adheres to practical design rules and remains deployable within the constraints of the target Nexys A7-100T FPGA. It serves as the most effective and scalable implementation for real hardware integration.

\subsection{BNN vs CNN Comparison Results}

We trained two neural network models for handwritten digit classification on the MNIST dataset as a binarized fully connected network and standard CNN. Both models are implemented in TensorFlow with fixed random seeds and tested under identical conditions. While both models process the same inputs, they differ in architecture, computational cost, and hardware suitability. The CNN uses two convolution layers (3$\times$3 filters with 32 and 64 channels), each followed by 2$\times$2 max pooling, a dense layer of 128 units with ReLU and dropout, and a softmax output. This architecture extracts spatial features but requires floating-point multiplications, non-linear activations, and large buffers. The BNN uses two hidden layers of 128 and 64 binary neurons and a binarized output layer, relying on XNOR-popcount operations and thresholding without multipliers or biases.

The CNN achieved 99.31\% accuracy while the BNN reached 87.97\% on the MNIST test set, reflecting the CNN’s superior spatial feature extraction. CPU inference latency over 100 runs averaged 0.176\,ms for the BNN and 0.213\,ms for the CNN, making the BNN approximately 17\% faster. Table~\ref{tab:cpu_latency} summarizes the statistical results.

\begin{table}[htbp]
\caption{CPU inference latency statistics across 100 runs.}\label{tab:cpu_latency}
\begin{center}
\begin{tabular}{ccccc}
\hline
\textbf{Model} & \textbf{Mean (ms)} & \textbf{Min (ms)} & \textbf{Max (ms)} & \textbf{Std Dev (ms)} \\
\hline
BNN & 0.176 & 0.149 & 0.276 & 0.022 \\
CNN & 0.213 & 0.191 & 0.293 & 0.016 \\
\hline
\end{tabular}
\end{center}
\end{table}

\begin{figure}[htbp]
\centerline{\includegraphics[width=\columnwidth]{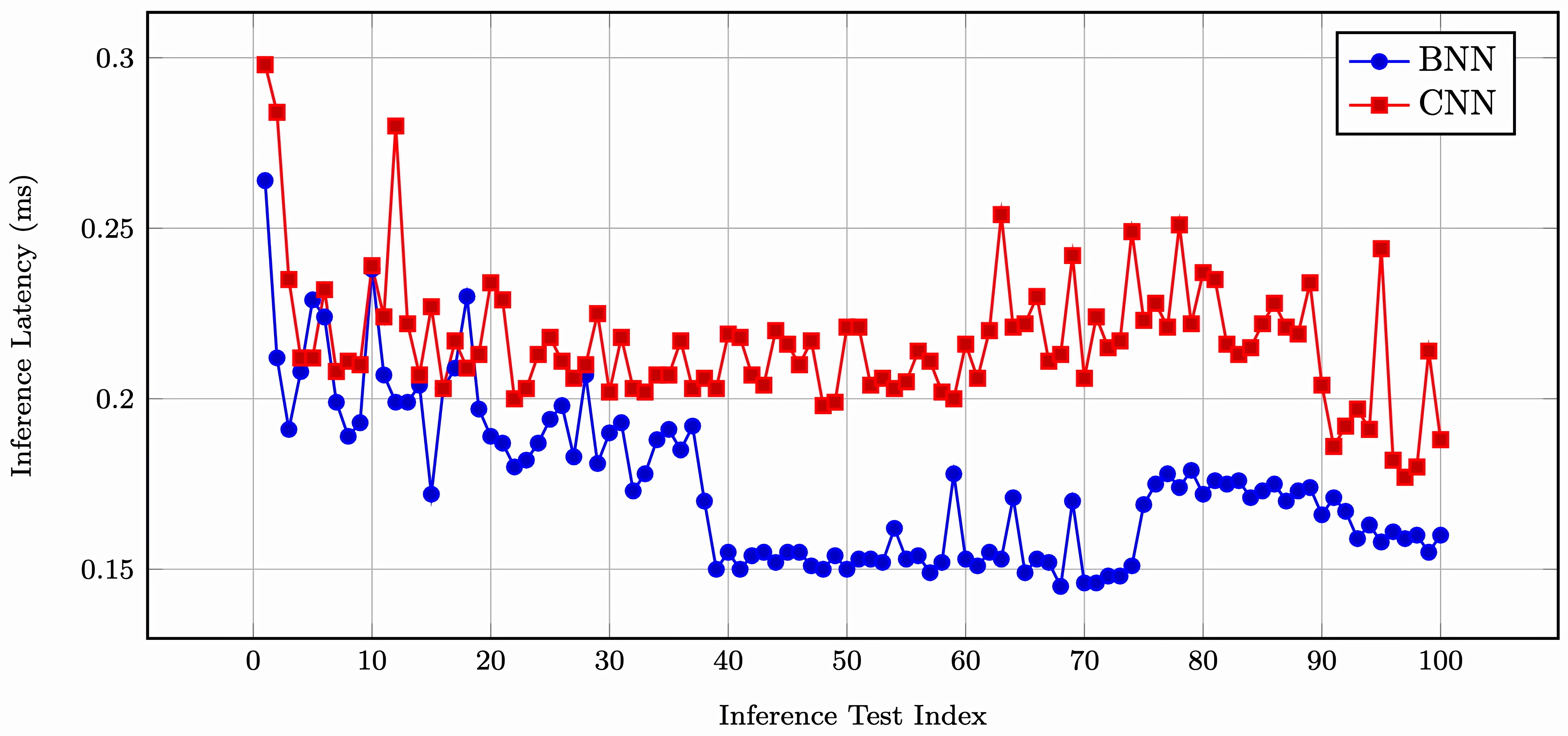}}
\caption{Inference latency on CPU for BNN and CNN models across 100 runs.}\label{figure:latency_plot}
\end{figure}

We also provide the run-by-run inference latency distribution in Fig.~\ref{figure:latency_plot}. As can be seen in this figure, the BNN consistently achieves lower inference latency compared to the CNN across all 100 runs. Moreover, it exhibits more stable behavior with smaller fluctuations and fewer performance spikes. This suggests that the BNN offers not only improved average inference speed but also more consistent runtime behavior. This is especially useful in real-time or resource-constrained applications.

The BNN model file size was 1.4\,MB versus 2.7\,MB for the CNN, due to 1-bit weights and a simpler structure. Training time was 15 seconds for the BNN (15 epochs) and 71 seconds for the CNN (10 epochs), as convolution operations, dropout, activations, and floating-point arithmetic make CNN training more computationally demanding. While the CNN requires fewer epochs to converge thanks to its feature extraction, the BNN’s faster training enables quicker iteration for hyperparameter tuning or retraining.

The BNN is well-suited for FPGAs, with bit-wise operations that map to LUTs, no need for DSP blocks, and a structure that is easy to parallelize and control. The CNN’s floating-point operations, DSP reliance, and large buffers require complex memory management and make timing closure harder. While more accurate, the CNN is less practical for efficient FPGA deployment.

\subsection{Platform Comparison Results}

This section evaluates the performance of the proposed FPGA-based BNN implementation against alternative platforms, focusing on power, latency, and efficiency. All comparisons use the 64-parallel configuration with BRAM on the Nexys A7-100T FPGA. 

\subsubsection{Comparison with ASIC}

Since direct ASIC implementation of the proposed BNN was beyond the scope of this study, an estimate-based comparison was performed using YodaNN, a low-power CNN accelerator with binary weights \cite{andri2016}. While YodaNN is optimized for convolutional layers, its binary-weighted architecture makes it a relevant ASIC reference. The YodaNN model shares key traits with this BNN, including binary weight representation and compact architecture. It operates at up to 480\,MHz and achieves a peak throughput of 1.5\,TOp/s at 1.2\,V, with core power consumption reported as 895\,$\mu$W at 0.6\,V. In terms of latency, the FPGA implementation achieves 0.0178\,ms per image at 80\,MHz, while YodaNN reports 7.5\,ms for a comparable three-layer binary model on CIFAR-10. Although YodaNN processes more complex inputs with convolution, this highlights the FPGA’s strength in low-latency inference.

For power, the FPGA’s total consumption is approximately 0.617\,W at 80\,MHz, including static and dynamic components. YodaNN’s inference power is inferred from its reported throughput and energy efficiency as
\[
P_{\text{ASIC}} \approx \frac{20.1\,\text{GOp/s}}{59.2\,\text{TOp/s/W}} = 0.00034\,\text{W}
\]
reflecting dynamic power during active computation only. The FPGA consumes about 11.0\,$\mu$J per inference, while YodaNN reports 2.6\,$\mu$J.

The YodaNN ASIC is estimated to cost \$5-\$10 per unit in mass production, though with high NRE costs and no post-fabrication flexibility. In contrast, the Nexys A7-100T FPGA development board costs approximately \$150 per unit, which is significantly higher than ASICs in volume but requires no fabrication setup. FPGAs provide full reconfigurability, allowing modifications to layer counts, neuron numbers, or dataflow without hardware redesign. While ASICs like YodaNN are optimized for energy and area efficiency, FPGAs offer adaptability and rapid iteration, making them more suitable for research, development, and prototyping.

\subsubsection{Comparison with CPU and GPU}

We run CPU and GPU benchmarks directly on the trained BNN model using Google Colab’s cloud resources. The CPU was an Intel Xeon @ 2.20\,GHz (2 logical cores), and the GPU was an NVIDIA Tesla T4 (70\,W TDP, 16\,GB GDDR6). Exact thermal and power metrics were unavailable due to the virtualized environment. To ensure fair timing, warm-up runs were performed and overhead minimized. Batch sizes from 1 to 10000 were tested. The results, tabulated in Table~\ref{table:cpu_gpu_results}, highlight latency and scalability results. Here, the mean, per image, and std dev inference times are given in ms.

\begin{table}[htbp]
\caption{Inference benchmark results across batch sizes.}\label{table:cpu_gpu_results}
\begin{center}
\begin{tabular}{rcrrr}
\hline
\textbf{Batch Size} & \textbf{Device} & \textbf{Mean} & \textbf{Per Image} & \textbf{Std Dev} \\
\hline
1 & CPU & 1.60 & 1.60000& 0.29 \\
1 & GPU & 0.82 & 0.82000 & 0.06 \\
10 & CPU & 1.01 & 0.10000 & 0.14 \\
10 & GPU & 0.87 & 0.08700 & 0.09 \\
100 & CPU & 1.75 & 0.01700 & 0.15 \\
100 & GPU & 1.22 & 0.01200 & 0.13 \\
1000 & CPU & 6.93 & 0.00690 & 1.13 \\
1000 & GPU & 0.86 & 0.00086 & 0.08 \\
10000 & CPU & 63.02 & 0.00630 & 15.47 \\
10000 & GPU & 1.58 & 0.00016 & 1.08 \\
\hline
\end{tabular}
\end{center}
\end{table}

As can be seen in Table~\ref{table:cpu_gpu_results}, GPU inference consistently outperformed CPU at higher batch sizes, with the GPU achieving a per-image latency of 0.16\,$\mu$s at batch 10{,}000, compared to 6.3\,$\mu$s on CPU. The GPU’s performance scaled well with batch size due to Tensor Core acceleration and memory-level parallelism, making it ideal for high-throughput inference. The CPU scaled reasonably up to batch 1000 but offered diminishing returns beyond, limited by parallelism and bandwidth. In comparison, the FPGA achieved 17.8\,$\mu$s per image (64-parallel at 80\,MHz), outperforming CPU in real-time, low-batch scenarios while consuming significantly less power (0.617\,W) than the GPU’s 70\,W. Unlike CPU and GPU, the FPGA provides deterministic timing and efficient resource usage, making it more suitable for embedded or latency-critical applications.

\subsubsection{Suitability of FPGAs for BNN Deployment}

Based on the experimental results and platform comparisons, FPGA stands out as the most practical platform for the proposed BNN architecture in this study. It combines fast inference, low power consumption, and reconfigurability without the high power of GPUs or the inflexibility and cost of ASICs. The final FPGA design achieved a per-image latency of 0.0178\,ms, significantly faster than CPU (1.60\,ms at batch size 1) and only slower than GPU at very large batch sizes. Its timing is deterministic, making it ideal for real-time systems where consistent response is critical, which is something CPUs and GPUs cannot always guarantee. The design consumes about 0.617\,W, far below the GPU’s 70\,W, supporting efficient embedded deployment with minimal cooling needs. All logic fit comfortably on the Nexys A7-100T board, using 26\% of LUTs and nearly all BRAM. Unlike ASICs, the FPGA can be reprogrammed to accommodate architectural changes or feature updates without hardware redesign. The board cost (about \$150) provides cost-effective development compared to GPUs (\$400–\$900), and while ASICs offer lower per-unit cost at scale (\$5–\$10), they include substantial NRE expenses. In summary, FPGA provides an optimal balance of speed, efficiency, flexibility, and development cost for this BNN and similar applications.

\section{Conclusions and Future Work}

This study presents the design, implementation, and evaluation of a BNN architecture for real-time handwritten digit recognition on FPGA. We can summarize the advantages of BNNs as follows. Since all weights and activations are stored using a single bit (-1 or +1), the overall model size can be reduced by a factor of 32 compared to networks that use 32-bit floating-point parameters. This allows much smaller memory usage, which is particularly useful in embedded systems. BNNs replace multiplication heavy operations with bitwise XNOR and popcount. These operations are faster and simpler to be implemented in hardware. As a result, inference latency significantly reduces, especially when compared to traditional floating-point neural networks. Binary operations use less switching and need fewer memory accesses, which keep power consumption at lower levels. This makes BNNs a good fit for situations where saving energy matters like in battery powered devices or low-power systems that need to run in real-time. However, these advantages come with some trade-offs. Binarization introduces approximation error by replacing floating-point values with binary ones, which lowers accuracy. BNNs are more sensitive to settings like learning rate, batch size, and weight initialization due to aggressive quantization. Hence, training usually requires more careful tuning to work well. Many standard layers and functions used in deep learning do not work well with binary values, which can reduce design flexibility or require custom solutions.

The proposed BNN architecture in this study is developed in two stages as Python-based training using TensorFlow and Larq, and Verilog-based inference system targeting the Nexys A7-100T. We verified the design through simulation, synthesis, implementation, and bitstream generation, confirming readiness for on-device usage. The final system achieved 84\% accuracy on the MNIST test set in behavioral simulation, closely matching software results on PC. It supports configurable parallelism up to 128 neurons per cycle and achieves 0.0178\,ms latency at 0.617\,W power consumption. Dual-port BRAMs enable parallel weight access, while thresholds folded from batch normalization are stored in LUT-based ROMs to minimize BRAM usage. A centralized FSM manages all inference stages. The Verilog design is validated through module-level testing and waveform inspection. Despite these results, several architectural limitations remains. The FSM is hardcoded for each individual layer, preventing support for variable-length networks or fully parameterized inference logic. In addition, the design focuses exclusively on fully connected layers and did not implement convolutional architectures, which are better suited for spatial feature extraction in image classification tasks. We plan to extend the proposed design implementation to SystemVerilog to support flexible, array-based parametrization of layer structures. This will enable dynamic FSM control based on a global array of neuron counts, eliminating the need for duplicated states. Additional enhancements include extending the architecture to support binary convolutional layers through the integration of line buffers, kernel sliding windows, and weight reuse patterns. To improve flexibility, future designs may incorporate SRAM-based weight storage, enabling runtime loading of model parameters without requiring resynthesis. Support for external image input, such as from a UART interface or low-resolution camera module, will enable real-world deployment, while UART-based output can provide digit predictions to external systems or microcontrollers. 

\bibliographystyle{unsrt} 
\bibliography{references}

\end{document}